\documentclass[twocolumn,pra,showpacs,nofootinbib,superscriptaddress]{revtex4}
\usepackage{amssymb}

\usepackage{dcolumn,bm,graphicx,amsmath}

\begin{document}

\preprint{UNR Jan 2005-\today }
\title{Quantum computing with magnetic atoms in optical lattices of reduced
periodicity}
\author{Boris Ravaine}
\affiliation{Department of Physics, University of Nevada, Reno,
Nevada 89557}
\author{Andrei Derevianko}
\affiliation{Department of Physics, University of Nevada, Reno,
Nevada 89557}
\author{P.R. Berman}
\affiliation{Michigan Center for Theoretical Physics, FOCUS
Center, and Department of Physics, University of Michigan, Ann
Arbor, Michigan 48109}
\date{\today}

\begin{abstract}
We investigate the feasibility of combining Raman optical lattices
with a quantum computing architecture based on lattice-confined
magnetically interacting neutral atoms. A particular advantage of
the standing Raman field lattices comes from reduced interatomic
separations leading to increased interatomic interactions and
improved multi-qubit gate
performance. Specifically, we analyze a $J=3/2$ Zeeman system placed in $%
\sigma _{+}-\sigma _{-}$ Raman fields which exhibit $\lambda /4$
periodicity. We find that the resulting CNOT gate operations times
are in the order of millisecond. We also investigate motional and
magnetic-field induced decoherences specific to the proposed
architecture.
\end{abstract}

 \maketitle

Controlled interactions between qubits is the key to practical
realization of quantum multi-qubit gates. The strength of the
interaction determines how fast the gate operations are performed.
In Ref.~\cite{DerCan04}, a quantum computing scheme based on
magnetically interacting atoms held in optical lattice was
proposed. Since the interaction between magnetic dipoles separated
by a distance $R$ scales as $1/R^{3}$, it is beneficial to reduce
the distance between the atoms. In traditional optical lattices,
created by two interfering counter-propagating laser fields of
wavelength $\lambda $, the interatomic distance is $\lambda /2$.
The estimates~\cite{DerCan04} show that a resulting CNOT-gate
performance time $\tau _{\mathrm{CNOT}}$ ranges from
$10^{-2}\,\mathrm{sec}$ for alkalis to $10^{-4}\,\mathrm{sec}$ for
complex open-shell atoms with large magnetic moments, such as
dysprosium.

Recently optical lattices of reduced, $\lambda
/2^{n}\,(n=1,2,3,\ldots )$, periodicity have been
proposed~\cite{DubBer02a,DubBer02b} and are under experimental
investigation~\cite{ZhaMorBer05}. Here we evaluate a feasibility
of combining such lattices with the quantum computing scheme of
Ref.~\cite{DerCan04}. Compared to the conventional $\lambda /2$
lattices, such a combination could potentially yield a factor of
$2^{3(n-1)}$ improvement in the gate performance time.
In this paper we analyze the case of a $\lambda /4$ lattice.

The paper is organized as follows. In Section~\ref{Sec:QC}, we
review the
relevant features of the quantum computing architecture with magnetic atoms~%
\cite{DerCan04}. In Section~\ref{Sec:lamO8}, we derive optical
potentials for a particular case of $J=3/2$ atoms and demonstrate
their $\lambda/4$ periodicity. In Section~\ref{Sec:QCops}, we
describe operation of our proposed quantum computing scheme.
Finally, in Section~\ref{Sec:decoh} we address important issues of
motional and magnetic-noise induced decoherences.

\section{Quantum computing with magnetic atoms}

\label{Sec:QC} In Ref.~\cite{DerCan04}, a scalable quantum
computing architecture was proposed. The architecture utilizes
magnetic interaction of complex open-shell atoms confined to the
nodes of an optical lattice. The lattice is placed in a high
gradient magnetic field and the resultant Zeeman sublevels define
qubit states. Microwave pulses tuned to space-dependent resonant
frequencies are used for individual addressing. Nearest neighbor
magnetic-dipolar atomic interactions allow for the implementation
of a quantum controlled NOT gate. For certain atoms, the resulting
single-qubit gate operation times are on the order of
microseconds, while the two-qubit operations require milliseconds.
These times are much faster than the anticipated decoherence
times.

While the magnetic interaction is weak (so the gate operations are
relatively slow), it is the goal of this paper to investigate a
potential route to strengthening inter-atomic interactions by
bringing atoms closer in optical lattices of reduced periodicity.
The proposed architecture offers several distinct advantages. For
example, compared to the popular scheme with Rydberg
gates~\cite{JakCirZol00}, the advantages are: (i) individual
addressing of atoms with \emph{unfocused} beams of microwave
radiation, (ii) coherent ``always-on'' magnetic-dipolar
interactions between the atoms, and (iii) substantial decoupling
of the motional and inner degrees of freedom.

Before proceeding further, it is worth remembering the following
order-of-magnitude values relevant to the architecture of Ref.~\cite%
{DerCan04}: light-shifts and the depth of the optical wells are
about 1 MHz, typical Zeeman splitings are 1 GHz, and the
difference in the resonant Zeeman frequency for two neighboring
wells is about 1 kHz.

\section{Optical lattices of reduced periodicity}

\label{Sec:lamO8} Optical lattices of reduced periodicity have
been proposed in Refs.~\cite{DubBer02a,DubBer02b}. In this section
we review the underlying laser field-atom configuration and
formalism, and then specialize the general formalism of
Ref.~\cite{DubBer02b} to the $J=3/2$ atomic Zeeman manifold.

\begin{figure}[th]
\begin{center}
\includegraphics[scale=0.5]{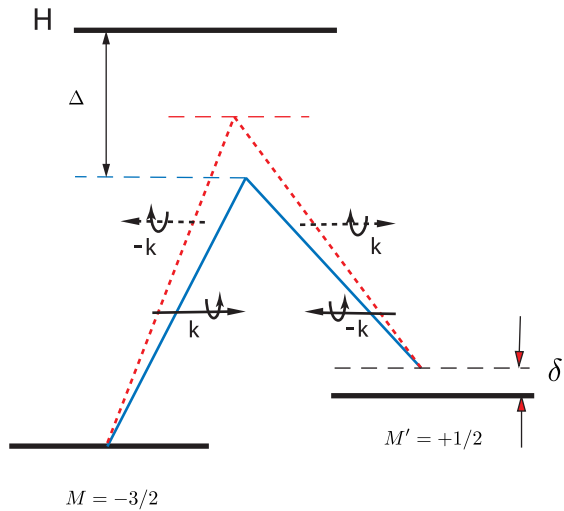}
\end{center}
\caption{The standing wave Raman atom-field configuration is
composed of
four laser fields. For $J_{g}=3/2$ the optically coupled states are either $%
M=-3/2,1/2$ or $M=-1/2,3/2$ and the lower-manifold splitting is
due to the Zeeman interaction. } \label{Fig:RamanSetup}
\end{figure}

In the atom-field geometry of Ref.~\cite{DubBer02b}, a neutral
atom interacts with four laser beams of equal intensities arranged
as two counter-propagating Raman pairs, see
Fig.~\ref{Fig:RamanSetup}.
The carrier frequencies of the pairs are denoted as $\left( \Omega
_{1}\text{
and }\Omega _{2}\right) $ and $\left( \Omega _{1}^{\prime }\text{ and }%
\Omega _{2}^{\prime }\right) $. The lasers are off-resonant with
the upper manifold $H$; neglecting a small difference in detunings
from the $H$ state for the two pairs, we use a single value for
the detuning $\Delta $, although it is this difference in detuning
that allows us to neglect
interference (modulated Stark shifts) between fields having frequences $%
\Omega _{1}$ and $\Omega _{1}^{\prime }$ (or $\Omega _{2}$ and
$\Omega _{2}^{\prime }$) that would give rise to the conventional
$\lambda /2$ periodicity of the optical lattice
\cite{BerRaiZha05,MalBer06}.
Each pair drives a two-photon transition between the substates $g$ and $%
g^{\prime }$ of the ground-state manifold. As shown in the figure,
the complementary fields $\Omega _{i}^{\prime }$ of each pair
could be detuned from resonance by $\delta =\Omega _{i}^{\prime
}-\Omega _{i}-\omega _{M^{\prime }M}$, where $\omega _{M^{\prime
}M}$ is the splitting between the ground state sublevels. In
addition to Raman-induced interactions each sublevel experiences
position-independent light-shifts due to the interaction with the
four individual laser fields.

We are interested in solving the Schr\"{o}dinger equation for the
described Raman atom-field geometry. At first we neglect atomic
center-of-mass (C.M.) motion and obtain solutions with the optical
Hamiltonian $H_{\mathrm{opt}}$ which incorporates internal atomic
Hamiltonian and the four atom-laser
interactions. We will return to the question of C.M. motion in Section~\ref%
{Sec:decoh}. Solving the time-dependent Schr\"{o}dinger equation,
\begin{equation}
i\frac{\partial }{\partial t}\tilde{\phi}\left( \xi ,z,t\right) =H_{\mathrm{%
opt}}\left( \xi ,z,t\right) \tilde{\phi}\left( \xi ,z,t\right) \,,
\label{Eq:TDSEDressed}
\end{equation}%
provides dressed atomic states $\tilde{\phi}(\xi ,z,t)$, where
$\xi $ and $z$ encapsulate internal and external (C.M.) degrees of
freedom, respectively.
Below we outline a method of solving the above equation developed in Ref.~%
\cite{DubBer02b}. To solve Eq.(\ref{Eq:TDSEDressed}), we
adiabatically eliminate the excited state and expand the dressed
states $\tilde{\phi}(\xi ,z,t)$ in terms of atomic stationary
states of the lower manifold. As a result one arrives at a system
of first-order linear differential equations for the amplitudes of
the ground state manifold. The RHS of the equations can be
expresses as a matrix multiplied by a vector of ground manifold
amplitudes. We denote this matrix $W$, it is easily reconstructed
from explicit expressions given by Eq.(17) of
Ref.~\cite{DubBer02b}. Diagonalization of the matrix $W$ produces
a set of position-dependent optical potentials $U_{i}(z)$ and
eigenvectors that define the dressed
states $\tilde{\phi}_{i}\left( \xi ,z,t\right) $. In Section~\ref{Sec:QCops}%
, we define qubit states in terms of these dressed states. Each
dressed state has a characteristic time-dependence
\begin{equation}
\tilde{\phi}_{i}\left( \xi ,z,t\right) =\phi _{i}\left( \xi
,z,t\right) e^{-iU_{i}(z)t}\,,  \label{Eq:untildedPhiDef}
\end{equation}%
the position-dependent phase leading to an optical force $-\nabla
U_{i}(z)$ acting on the atom.

Having reviewed the general standing-wave Raman setup and the
accompanying formalism, we specialize our discussion to an atom
with the total angular momenta of $J_{g}=3/2$ for the lower
manifold and $J_{h}=5/2$ for the upper manifold. A practically
relevant example is the metastable $3p_{3/2}$ state
of Al atom. It has been used in a matter-wave deposition experiment~\cite%
{McGGilLee95}, where the atoms were cooled on the closed transition to the $%
3d_{5/2}$ state ($\lambda =309\,\mathrm{nm}$). We estimate the
lifetime of the $3p_{3/2}$ state to be in the order of $10^{4}$
seconds, much longer than the anticipated
decoherence/loading/cooling time scales. A number of other
open-shell atoms have $J=3/2$ ground states as well.

In the B-field required for addressing individual atoms, the
$J=3/2$ manifold splits into four Zeeman levels. For Al, the Lande
factor is 4/3 and the Zeeman ladder in the B-field $B_{0}$ is
given by
\begin{equation*}
E_{M}=\frac{4}{3}\mu _{B}B_{0}M\,.
\end{equation*}%
The $\sigma ^{+}$-$\sigma ^{-}$ Raman fields couple only $M=-3/2$
to $M=1/2$ and $M=-1/2$ to $M=3/2$ and effectively we deal with a
pair of independent two-level subsystems. Moreover, while
constructing the matrices $W$, we find that they are equivalent;
the only difference between the subsystems comes from the fact
that in the B-field the level $M=-3/2$ is below $M=1/2$, while
$M=3/2$ is above $M=1/2$. For zero Raman detuning $\delta $ this
symmetry leads to identical optical potentials for the two
sub-systems. For non-zero detunings, the potentials are related to
each other by changing $\delta \rightarrow -\delta $. The
resulting \emph{four} optical potentials may be parameterized as
\begin{eqnarray}
U_{\pm }^{-3/2,1/2}(z) &=&-\frac{\alpha }{3}\pm
\frac{1}{30}\sqrt{(\alpha
+15\delta )^{2}+3\alpha ^{2}\cos ^{2}(2kz)}\,,  \notag \\
U_{\pm }^{3/2,-1/2}(z) &=&-\frac{\alpha }{3}\pm
\frac{1}{30}\sqrt{(\alpha -15\delta )^{2}+3\alpha ^{2}\cos
^{2}(2kz)},  \label{pot}
\end{eqnarray}%
where the superscripts in $U_{\pm }^{M,M^{\prime }}(z)$ specify
optically
coupled Zeeman pairs. Here we introduce the reduced dynamic polarizability $%
\alpha =\chi ^{2}/\Delta $, with the Rabi frequency $\chi
=-(1/2)\langle H||D||G\rangle E/\sqrt{3}$ defined in terms of the
reduced dipole matrix element and the (equal) strengths of
individual laser E-fields.

The derived optical potentials and the corresponding dressed
states depend on the adjustable detuning $\delta $, see
Fig.~\ref{Fig:RamanSetup}. Because of the $\delta \leftarrow
-\delta $ mapping, for $\delta =0$ the optical potentials for the
two sets of Zeeman levels coincide. However, the resulting $U_{+}$
and $U_{-}$ potentials are energetically shifted with respect to
each other and this energy gap might lead to a non-unform loading
of the lattice. Below we optimize the choice of the detuning
$\delta $.

\begin{figure}[th]
\begin{center}
\includegraphics[scale=0.75]{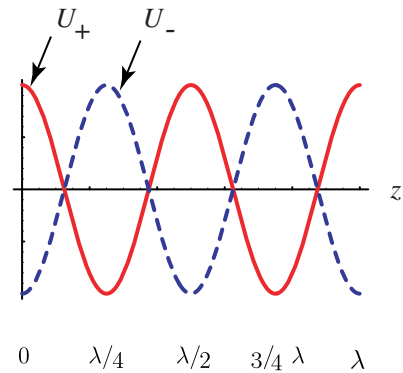}
\end{center}
\caption{Optical potentials, Eq.(\protect\ref{Eq:UpmOptimal}), for
the optimal choice of Raman detuning $\protect\delta $. While each
individual potential $U_{+}(z)$ and $U_{-}(z)$ has a
$\protect\lambda /2$ periodicity, the minima of the potentials
interleave, with the resulting distance between trapped atoms
being $\protect\lambda /4$. } \label{Fig:OptPot}
\end{figure}

First we focus on the $-3/2,+1/2$ optically coupled Zeeman pair. A
particular choice of $\delta =-\alpha /15$ leads to a pair of
potentials,
\begin{equation}
U_{\pm }(z)=-\frac{\alpha }{3}\pm \frac{\alpha }{10\sqrt{3}}\cos
\left( 2kz\right) \,,  \label{Eq:UpmOptimal}
\end{equation}%
that are energetically equivalent (see Fig.~\ref{Fig:OptPot}). As
an additional benefit, at this value of $\delta $ the potentials
have the largest intersite barriers.
In going from Eq. (\ref{pot}) to (\ref%
{Eq:UpmOptimal}), we have taken $\sqrt{\cos ^{2}(2kz)}=\cos \left(
2kz\right) $ rather than $\left\vert \cos \left( 2kz\right)
\right\vert $ to avoid important non-adiabatic coupling between
the potentials that would occur had we taken the absolute value
~\cite{BerRaiZha05,MalBer06}. Qualitatively, the detuning $\delta
=-\alpha /15$ is chosen to compensate for the difference in the
light shifts for the two sub-levels. As shown in
Fig.~\ref{Fig:OptPot}, the minima of each individual potential are
separated by $\lambda /2$. In addition, the potentials are shifted
with respect to each other. This produces potential minima
separated by $\lambda /4$. In other words, if the
lattice is properly loaded, the distance between the neighboring atoms is $%
\lambda /4$. Compared to the conventional $\lambda /2$ lattices,
the use of the described Raman configurations increases the atomic
dipolar interactions by a factor of $2^{3}$. The eigenstates of
the optical Hamiltonian (omitting the potential dependence, c.f.
Eq.(\ref{Eq:untildedPhiDef})) are
\begin{align}
\phi _{\pm }\left( \xi ,z,t\right) &
=\frac{1}{\sqrt{2}}(\left\vert
-3/2\right\rangle e^{-i(E_{-3/2}-\delta /2)t}\mp   \notag \\
& \left\vert +1/2\right\rangle e^{-i(E_{1/2}+\delta /2)t})
\label{Eq:phiOpt}
\end{align}%
Of course, the separation between adjacent potential minima in
conventional \textit{lin}$\perp $\textit{lin} lattices is also
$\lambda /4$, but each adjacent minima contains atoms in a given
magnetic substate and is not suitable for the computing scheme
described in this paper.

The problem with the above choice of detuning $\delta =-\alpha
/15$ is that
it optimizes only the potentials for the $-3/2,+1/2$ pair. For the other, $%
+3/2,-1/2$, pair the optimal choice would be $\delta ^{\prime }=+\alpha /15$%
. If we keep the $\delta =-\alpha /15$ detuning, the height of the
intersite barriers for the $+3/2,-1/2$ pair would be just $\sim
20\%$ of the optimal values and in addition the corresponding
$U_{+}(z)$ and $U_{-}(z)$ potentials would be energetically
separated. To rectify this problem, we envision adding another
independent set of four laser beams. To distinguish between the
two sets, we will use primed quantities for this second set. This
second quadruplet would be Raman resonant for the $+3/2,-1/2$
coupled sublevels. The required detuning is $\delta ^{\prime
}=\alpha /15=-\delta $. We further require that the relative
phases of the laser fields for the two quads are adjusted so that
both resulting sets of $U_{+}/U_{-}$ potentials coincide, i.e.,
$U_{\pm }^{\prime }(z)\equiv U_{\pm }(z)$. The eigenstates are
\begin{align}
\phi _{\pm }^{\prime }\left( \xi ,z,t\right) & =\frac{1}{\sqrt{2}}%
(\left\vert 3/2\right\rangle e^{-i(E_{3/2}-\delta /2)t}\mp   \notag \\
& \left\vert -1/2\right\rangle e^{-i(E_{-1/2}+\delta /2)t})\,.
\label{Eq:phiOptPrime}
\end{align}%
We assume that the dominant trapping is produced by the
(two-photon) resonant fields, but detailed calculations are needed
to confirm this assumption.

Notice that in the B-field gradient the Raman detuning $\delta $
would change across the lattice sites. However, the change ($\sim
1\,\mathrm{kHz}$
increment per site) is small compared to the typical light-shifts, $\sim 1\,%
\mathrm{MHz},$ implying that the resonance conditions $\delta =\pm
\alpha /15 $ is little changed over the entire sample.

To summarize results of the discussion so far, we have designed a
confinement scheme for $J=3/2$ atoms in magnetic fields. The atoms
are separated by a distance of $\lambda/4$, improving the
performance of multi-atom gates over the conventional $\lambda/2$
schemes. In the following
section we address operations of a quantum computer based on the described $%
\lambda/4$ lattice.

\section{Operation}

\label{Sec:QCops} First we define the qubit states in terms of the
eigenstates, Eqs.~(\ref{Eq:phiOpt},\ref{Eq:phiOptPrime}), of the
optical Hamiltonian. We assume that the atoms are in the
Lamb-Dicke confinement
regime, and the atoms are trapped at the minima of the $U_{+}(z)$ and $%
U_{-}(z)$ potentials (Fig.~\ref{Fig:OptPot}) with 1:1 occupation
ratio. The definition of the qubit depends on whether the atom is
trapped at $U_{+}(z)$ or $U_{-}(z)$ minima. For the $U_{+}(z)$
wells,
\begin{equation*}
z_{n}^{\left( +\right) }=\frac{\lambda }{2}n,\left\{
\begin{array}{c}
\left\vert 1\right\rangle =\phi _{+} \\
\left\vert 0\right\rangle =\phi _{+}^{\prime }\,,%
\end{array}%
\right.
\end{equation*}%
while for the $U_{-}(z)$ wells
\begin{equation*}
z_{n}^{\left( -\right) }=\frac{\lambda }{4}+z_{n}^{\left( +\right)
},\left\{
\begin{array}{c}
\left\vert 1\right\rangle =\phi _{-} \\
\left\vert 0\right\rangle =\phi _{-}^{\prime }%
\end{array}%
\,,\right.
\end{equation*}%
where $n$ is an integer. To individually address the qubits, we
introduce a B-field gradient and use pulses of microwave radiation
of various duration to execute one-qubit operations. However,
compared to Ref.~\cite{DerCan04}, here we deal with the dressed
states. Describing dynamics of the system requires certain care.
For concreteness, we focus on an atom in the $U_{+}$ well, but the
conclusions will apply equally to the $U_{-}$ wells. The total
Hamiltonian including interaction $V_{M}$ with
circularly-polarized MW radiation of frequency $\omega $ reads
$H=H_{\mathrm{opt}}+V_{M}e^{-i\omega t}+V_{M}^{\dagger
}e^{+i\omega t}$. The duration of the drive should be chosen to
resolve different Zeeman transition frequencies between ground
state sublevels near the potential minima of a single well, as
well as transition frequencies between adjacent $U_{+}$ and
$U_{-}$ potential wells. These conditions can be satisfied easily,
as the change of the Zeeman frequency for two neighboring sites is
$\sim ~1\,\mathrm{kHz}$, while the light-shifts are in the order
of 1 MHz. Under such an assumption, the atomic
wavefunction can be expanded as $\Psi (t)=c(t)\tilde{\phi}_{+}+c^{\prime }(t)%
\tilde{\phi}_{+}^{\prime }$. Taking into account
Eq.(\ref{Eq:TDSEDressed}), we arrive at the system of coupled
equations for the expansion amplitudes,
\begin{eqnarray*}
i\dot{c} &=&\langle \phi _{+}|V_{M}e^{-i\omega t}+V_{M}^{\dagger
}e^{+i\omega t}|\phi _{+}^{\prime }\rangle c^{\prime }(t) \\
i\dot{c^{\prime }} &=&\langle \phi _{+}^{\prime }|V_{M}e^{-i\omega
t}+V_{M}^{\dagger }e^{+i\omega t}|\phi _{+}\rangle c(t)
\end{eqnarray*}%
While evaluating matrix elements we find that the resonant frequencies are $%
\omega _{\mathrm{res}}=\omega _{Z},\omega _{Z}\pm \delta $, where
$\omega _{Z}=\frac{4}{3}\mu _{B}B_{0}$. The resonance at $\omega
_{Z}$ corresponds to transitions between the $M=-1/2$ and $M=1/2$
sublevels, while those at$\
\omega _{Z}\pm \delta $ correspond to transitions between $M=-3/2$ and $%
M=-1/2$ or $M=1/2$ and $M=3/2$. Again, due to the
orders-of-magnitude difference in the Zeeman and the light-shift
energy scales, we may resolve these resonance frequencies and find
it convenient to work at $\omega _{Z}$. For the selected
transition, the above system of equations maps onto the problem of
a two-level system in an oscillating field. By varying the
duration of the MW pulses, we may execute arbitrary rotations in
the Hilbert space spanned by individual qubits.

Since the Zeeman splitting is position-dependent, only a single
qubit from the entire ensemble will respond to a pulse of a
certain frequency. The advantage of the proposed addressing scheme
is that there is no need to focus radiation on an individual atom.

Now we turn to another important ingredient of QC architectures:
multi-qubit gates. It is sufficient to consider operation of the
universal two-qubit CNOT gate~\cite{BarDeuEke95}. In the proposal
of Ref.~\cite{BarDeuEke95} the resonance frequency of the target
qubit should depend on the state of the control qubit. This leads
to a conditional quantum dynamics: the transition in the target
qubit occurs only if the control qubit is in a predefined state.
In our scheme, this dependence of the resonance frequency is due
to magnetic interactions between two neighboring atoms. Indeed, we
find that each qubit state posses a permanent magnetic-dipole
moment aligned with the z-axis: $\mu _{|1\rangle }=+2/3\mu _{B}$,
$\mu _{|0\rangle }=-2/3\mu _{B}$ (these values are independent of
the well). The generated magnetic field at the position of the
target qubit is $\delta B=2\alpha ^{2}/R^{3}\mu _{|1,0\rangle }$,
where $\alpha \approx 1/137$ is the fine-structure constant and
$R=\lambda /4$ is the distance between the qubits. Depending on
the state of the control qubit the generated field will either
increase or reduce the offset B-field $B_{0}(z)$ and modify the
Zeeman frequency. While performing the gate, one needs to resolve
the frequency difference
\begin{equation*}
\delta \omega _{\mathrm{CNOT}}=\frac{32}{9}\alpha ^{2}\frac{\mu _{B}^{2}}{%
R^{3}}\,.
\end{equation*}%
The minimum duration of the MW pulse is $\tau _{\mathrm{CNOT}}\sim
1/\delta \omega _{\mathrm{CNOT}}$. For our parameters, $\tau
_{\mathrm{CNOT}}\approx 10^{-3}\,\mathrm{s}$. The resulting
performance is competitive with other
quantum computing schemes such as nuclear magnetic resonance~\cite%
{LafKniCor02} ($\tau _{\mathrm{CNOT}}\sim 10^{-3}-10^{-2}$ s), and
controlled collisions~\cite{CalHinJak00} ($\tau
_{\mathrm{CNOT}}\sim 4\times 10^{-4}$ s).


\section{Motional and magnetic-noise induced decoherences}

\label{Sec:decoh} The gate operations must be much faster than
decoherence rates. A number of decoherence mechanisms present in
the current proposal have been analyzed in Ref.~\cite{DerCan04}.
For example, atoms may be lost due to an entanglement of the
internal and motional degrees of freedom during the NOT gate
operation. It was demonstrated~\cite{DerCan04} that the associated
excitation rates from the ground motional state are negligible.
The underlying reason is that the induced perturbation is
adiabatically slow: on the time scale of the NOT pulse, the atomic
C.M. undergoes many oscillations in the well. The same conclusion
holds for the present
proposal. However, the dressed (qubit) states were introduced in Sec.~\ref%
{Sec:lamO8} neglecting the C.M. motion and we need to additionally
consider coupling of dressed states due to the atomic motion.
Fortunately, as shown below, it is straightforward to demonstrate
that the qubit states are not coupled by the C.M. motion.

Another important source of the decoherence arises due to magnetic
noise. The qubit states are defined in terms of the Zeeman
sublevels sensitive to magnetic perturbations. In
Ref.~\cite{DerCan04}, a noise-induced dephasing has been
evaluated, and it has been shown that the decoherence rate can be
reduced with modest B-field shielding requirements. In the present
proposal, the implications of the magnetic noise can be more
severe: for dressed states, the relative phase (determined by the
Zeeman splitting) between the ``bare'' magnetic substates must
remain fixed. The noise perturbs the relative phase leading to
excitations of the motional quanta.

In Section II, the dressed states $\tilde{\phi}_{\pm }(\xi ,z,t)$
were found by assuming that the atom was localized at the position
$z$ and then by diagonalizing the position-dependent optical
matrix $W(z)$. Taking into account that for a fixed $z$, the
$\tilde{\phi}_{\pm }(\xi ,z,t)$ form a complete basis set, we may
expand the exact wave function as
\begin{equation}
\Psi (\xi ,z,t)=\chi _{-}(z,t)\tilde{\phi}_{-}(\xi ,z,t)+\chi _{+}(z,t)%
\tilde{\phi}_{+}(\xi ,z,t)\,,
\end{equation}%
where $\chi _{\pm }(z)$ are (generally coupled) C.M. wave
functions of interest. The Hamiltonian including the C.M. motion
is
\begin{equation}
H(\xi ,z,t)=T+H_{\mathrm{opt}}(\xi ,z,t)\,,
\end{equation}%
with $T=-\frac{1}{2M}\frac{\partial ^{2}}{\partial z^{2}}$ being
the kinetic energy operator for the C.M. motion. Using the
standard technique of projecting the Schr\"{o}dinger equation onto
the $\tilde{\phi}_{\pm }$ basis, we find
\begin{equation}
i\frac{\partial }{\partial t}\chi _{\pm }(z,t)=\sum_{p=\pm
}\langle \phi _{\pm }|T|\phi _{p}\chi _{p}\rangle _{\xi }+U_{\pm
}(z)\chi _{\pm }(z,t)\,,
\end{equation}%
where the inner product is with respect to the internal degrees of freedom $%
\xi $. The coupling between the two components $\chi _{+}$ and
$\chi _{-}$ arises in general due to the off-diagonal term in the
sum. In our case, the
dressed states $\phi _{\pm }(\xi ,t)$ do not depend on $z$, Eqs.(\ref%
{Eq:phiOpt}) and (\ref{Eq:phiOptPrime}), and we arrive at a simple
result
\begin{equation}
i\frac{\partial }{\partial t}\chi _{\pm
}(z,t)=-\frac{1}{2M}\frac{\partial ^{2}}{\partial z^{2}}\chi _{\pm
}(z,t)+U_{\pm }(z)\chi _{\pm }(z,t)\,. \label{sSchro}
\end{equation}%
This is a physically significant result: the atomic motion does
not lead to mixing of the qubit states.

At the minima of the potentials, the bottom of the potential well
can be approximated by a harmonic oscillator potential and we can
write $\chi _{\pm }(z,t)$ as a sum of the time-dependent
stationary-states of the harmonic oscillator $\chi _{n}$,
\begin{equation}
\chi _{\pm }(z,t)=\sum_{n}c_{\pm ,n}(t)\chi _{n}(z,t)\,.
\end{equation}%
The coefficients $c_{\pm ,n}$ depend on the temperature of the
qubit and the loading process; we assume that initially only the
ground motional states are occupied. We also require that the
potential is sufficiently deep so that the atoms do not tunnel
away, see Ref.~\cite{DerCan04} for estimates.

Now we can analyze the effect of time-dependent magnetic noise
$B(t)$ acting on the qubit. Let us take, for example, the qubit in
the state $\phi _{+}$
localized in the motional ground state $\chi _{0}$ of one of the minima of $%
U_{+}$ for $t<0$. For $t\geq 0$, the magnetic noise is turned on
and we look at the decoherence rate of the qubit. We consider the
loss mechanism as a two-step process. The magnetic field acts only
on the internal degree of freedom of the qubit causing a primary
transition $c_{+,0}\,\phi _{+}\,\chi _{0}(z)\rightarrow
c_{-,0}\,\phi _{-}\,\chi _{0}(z)$. Minima of $U_{+}(z)$ correspond
to maxima of $U_{-}(z)$ so the new state $\phi _{-}\,\chi _{0}(z)$
is motionally unstable. For example, one of the possible
transitions this state undergoes is $c_{-,0}\,\phi _{-}\,\chi
_{0}(z)\rightarrow c_{-,2}\,\phi _{-}\,\chi _{2}(z)$. The exact
details of this secondary transition are not important, since the
entire decoherence rate is determined by the rate of the primary
excitation. For example, characteristic time associated with the
$c_{-,0}\,\phi _{-}\,\chi
_{0}(z)\rightarrow c_{-,2}\,\phi _{-}\,\chi _{2}(z)$ route is $\tau _{2}=%
\frac{1}{2k}\sqrt{\frac{5\sqrt{3}M}{\alpha }}$. Our estimate for Al atom, $%
\lambda =309\,\mathrm{nm}$ and $\alpha \sim 10^{7}\,\mathrm{Hz}$ results in $%
\tau _{2}\sim 10^{-7}\,\mathrm{s}$.

The decoherence rate is determined by the primary excitation
$c_{+,0}\,\phi _{+}\,\chi _{0}\rightarrow c_{-,0}\,\phi _{-}\,\chi
_{0}$ and below we determine the probability associated with this
process. The required time evolution of the coefficient
$c_{-,0}(t)$ is computed by including the magnetic noise
perturbation $\mu B(t)\,|\phi _{-}\rangle \langle \phi
_{+}|\,+h.c.\,$ in the Schr\"{o}dinger equation
\begin{equation}
i\frac{\partial }{\partial t}c_{-,0}(t)=(U_{\mathrm{MAX}}-U_{\mathrm{MIN}%
})c_{-,0}(t)+\mu B(t)c_{+,0}(t)\,,  \label{c-eq}
\end{equation}%
where $\mu =\langle \phi
_{-}|\mathbf{\hat{\mu}}\frac{\mathbf{B}}{B}|\phi
_{+}\rangle $, and $U_{\mathrm{MAX}}-U_{\mathrm{MIN}}=\frac{\alpha }{5\sqrt{3%
}}$ is the energy difference between the minimum and the maximum of the $%
U_{+}$ and $U_{-}$ potentials, Eq(\ref{Eq:UpmOptimal}). The solution of Eq.~(%
\ref{c-eq}) is
\begin{equation}
c_{-,0}(t)=-ie^{-i\frac{\alpha }{5\sqrt{3}}t)}\int_{0}^{t}\mu
B(t_{1})c_{+,0}(t_{1})e^{i\frac{\alpha }{5\sqrt{3}}t_{1}}dt_{1}\,.
\end{equation}%
The magnetic noise is characterized by its autocorrelation
function
\begin{equation}
\langle B(t_{1})B^{\ast }(t_{1}+\tau )\rangle =\frac{1}{2\pi
}\int_{-\infty }^{+\infty }S(\omega )e^{i\omega \tau }d\tau \,,
\end{equation}%
where $S(\omega )$ is the frequency-dependent spectral density of
the noise. The autocorrelation function $\langle
c_{-,0}(t)c_{-,0}(t)\rangle $ represents the probability $p(t)$ of
excitation $\phi _{+}\,\chi _{0}\rightarrow \phi _{-}\,\chi _{0}$,
\begin{eqnarray}
p(t)=\int_{0}^{t}\int_{0}^{t}\!\!\! &dt_{1}&\!\!\!dt_{2}\,\mu
^{2}\;\;\;\;\;\; \\
&\langle B&\!\!\!(t_{1})c_{+,0}(t_{1})B^{\ast
}(t_{2})c_{+,0}^{\ast }(t_{2})e^{i\frac{\alpha }{\hbar
5\sqrt{3}}(t_{1}-t_{2})}\rangle \,.  \notag
\end{eqnarray}%
For time $t$ sufficiently short to guarantee that $|c_{-,0}(t)|\ll
|c_{+,0}(t)|,$ but longer than the correlation time of the
magnetic noise, we may approximate $c_{+,0}(t)\simeq 1$, and find
\begin{equation}
p(t)=\mu ^{2}\int_{0}^{t}S(-\frac{\alpha }{5\sqrt{3}})dt=\frac{1}{\tau _{1}}%
t,
\end{equation}%
with $\tau _{1}=\frac{1}{\mu ^{2}}S(\frac{\alpha
}{5\sqrt{3}})^{-1}$.

For coherence times on the order of 10 seconds the magnetic noise
should satisfy
\begin{equation}
\sqrt{S(\frac{\alpha }{5\sqrt{3}})}\ll 3\times 10^{-12}\,\mathrm{{T}/\sqrt{{%
Hz}}\,.}
\end{equation}%
This level can be attained with passive
shielding~\cite{BudKimDeM03}. Moreover, notice that the spectral
density of the noise is evaluated at a relatively high ($\sim
1\,\mathrm{MHz}$) frequency. This frequency is of the order of the
light shift induced by the lattice lasers. At such a high
frequency the magnetic noise is highly suppressed. For example,
for passive shielding, the characteristic cut-off frequency due to
induced currents is in the order of kHz~\cite{BudKimDeM03}.

To conclude, our analysis suggests that magnetic noise can be
controlled at an adequate level. Also the atomic motion does not
lead to entanglement of qubit states, defined as dressed atomic
Zeeman sublevels. Other sources of
decoherence common with the present proposal, were considered in Ref.~\cite%
{DerCan04} and we refer the reader to that paper for details.

\section{Conclusion}

We have outlined a method for increasing the relatively weak
magnetic interactions in which atoms are trapped in Raman optical
lattices having reduced periodicity. The reduced interatomic
distances lead to improved performance of multi-qubit gates. In
the particular case of the Al, $J=3/2$ Zeeman manifold, we
designed a $\lambda /4$ optical lattice and found that
universal two-qubit CNOT gate operations require times of approximately $%
10^{-3}$~s. These times are comparable to other quantum computing
schemes such as nuclear magnetic resonance~\cite{LafKniCor02} and
controlled collisions~\cite{CalHinJak00}. Moreover, the present
proposal offers scalability, individual qubit addressability with
\emph{unfocused} beams of microwave radiation, and coherent
\textquotedblleft always-on\textquotedblright\ interactions
between the qubits.

Analysis of Refs.~\cite{DubBer02a,DubBer02b} suggests that in
general, the standing-wave Raman fields should lead to
$\lambda/2^{n} \, (n=1,2,3,...)$ interatomic separations. For a
given number of atoms, such lattices should improve performance of
the original $\lambda/2$ quantum computing architecture of
Ref.~\cite{DerCan04} by an exponentially increasing factor of
$2^{3(n-1)}$. It remains to be seen if the present $\lambda/4$
proposal can be generalized to optical lattices of smaller
periodicity.

\acknowledgements We would like to thank D. Budker and D. DeMille
for discussions. A.D. would like to thank for the hospitality the
FOCUS center, where a part of this work has been completed. This
work was supported in part by the NSF Grant Nos. PHY-0354876,
PHY0244841, the NSF FOCUS Center Grant, and the Michigan Center
for Theoretical Physics.
\bibliographystyle{apsrev}

\end{document}